\documentclass[aps,showpacs,nofootinbib]{revtex4}

\usepackage{graphicx}

\begin{document}

\title{The form factor of the pion in ``point-form'' of relativistic
dynamics revisited} 

\author{A. Amghar}
\affiliation{Facult\'e des Sciences, Universit\'e de Boumerdes, 
35000 Boumerdes, Algeria}

\author{B. Desplanques} 
\email{desplanq@isn.in2p3.fr} 
\affiliation{Institut des Sciences Nucl\'eaires (UMR CNRS/IN2P3--UJF),
F-38026 Grenoble Cedex, France}

\author{L. Theu{\ss}l} 
\email{Lukas.Theussl@uv.es} 
\affiliation{Departamento de Fisica Teorica, Universidad de Valencia,
E-46100 Burjassot (Valencia), Spain}

\date{\today}

\begin{abstract}
The electromagnetic form factor of the pion is calculated in the 
``point-form'' of relativistic quantum mechanics using simple, 
phenomenological wave functions. It is found that the squared charge radius 
of the pion is predicted one order of magnitude larger than the experimental
value and the asymptotic behavior expected from QCD cannot be reproduced. 
The origin of these discrepancies is analyzed.  The present results 
confirm previous ones obtained from a theoretical model and call for major 
improvements in the implementation of the ``point-form'' approach.
\end{abstract} 

\pacs{13.40.Gp, 12.39.Ki, 14.40.Aq}

\maketitle

\section{Introduction}

\noindent
The pion charge form factor has been the object of many 
studies~\cite{Isgur:1984jm,Chung:1988mu,Cardarelli:1995dc,Roberts:1996hh,
Choi:1997iq,deMelo:1997cb,Maris:2000sk,Krutov:2002nu,Allen:1998hb}. 
The simplicity of the system, a quark -- anti-quark bound state, and the 
availability of experimental 
data~\cite{Bebek:1978pe,Amendolia:1986wj,Volmer:2000ek} make it
very attractive to test both physical ingredients and  methods employed 
in its calculation. In view of a strongly bound system (if one assumes 
constituent quark masses of the order of a few hundred MeV) and 
the large $Q^2$ at which the pion form factor has been measured, 
a relativistic calculation is most likely required. There are various 
approaches to implement relativity, ranging from field theory 
to relativistic quantum mechanics. Among the latter ones, calculations 
of the pion form factor have been done in instant and front 
forms~\cite{Isgur:1984jm,Chung:1988mu,Cardarelli:1995dc,Roberts:1996hh,Choi:1997iq,deMelo:1997cb,Maris:2000sk,Krutov:2002nu}. 
Only one has been performed in the less known 
``point-form''~\cite{Klink:1998}. A good agreement with experiment was 
claimed by the authors~\cite{Allen:1998hb}. 

In this ``point-form'' approach
the parameter determining the elastic form factor is the relative 
velocity of the initial and final states. Thus, in the  Breit frame, 
the form factor depends on the momentum transfer $Q^2$ only through 
the quantity, $v^2=Q^2/(4\,M^2+Q^2)$, where $M$ is the total mass 
of the system. This has the immediate consequence that the squared charge 
radius of the system
scales like $1/M^2$, and therefore tends to $\infty$ when $M$ goes to zero. 
Accordingly, the charge radius increases when the 
system becomes stronger bound, contrary to physical intuition. 
For the pion, this argument suggests that the
squared charge radius should be of the order 
of $3/(2\,M_{\pi}^2) \simeq 3\,{\rm fm}^2$, which exceeds the 
experimental value by almost one order of magnitude. In contrast 
to this result, the calculation of Ref.~\cite{Allen:1998hb} shows nice
agreement. Furthermore, these authors show that,   
in the case of an infinitely bound system, the form factor 
could scale like $1/Q^2$ for large momentum transfers, 
in agreement with the QCD expectation~\cite{Brodsky:1975vy,Lepage:1979zb}.
However, the general derivation of this result does not imply any assumption 
on the binding of the system. 

Motivated by the above observations,   
we re-examine in this letter the ``point-form'' calculation of the 
pion form factor. 
While doing so, we also have in mind the 
failure of the approach when applied to a system made of scalar 
constituents~\cite{Desplanques:2001zw,Amghar:2002jx}. 
The main goal of this study is a clarification of the above-mentioned 
peculiarities connected with recent implementations of the 
``point-form'' of relativistic Hamiltonian 
mechanics\footnote{As it has been observed in our recent 
work~\protect\cite{Desplanques:2001zw,Amghar:2002jx,Desplanques:2001ze} and
in Ref.~\protect\cite{Sokolov:1985jv}, this
implementation is not identical to the form proposed by 
Dirac~\protect\cite{Dirac:1949cp} in that it does not involve a quantization
performed on a hyperboloid. To emphasize this difference, we will put the
expression ``point-form'' between quotation marks throughout this paper.}.
A deeper understanding of this relatively unknown approach 
should certify that the method is on a safe ground. 
With this aim we will use simple wave functions (a Gaussian one as used in 
Ref.~\cite{Allen:1998hb} and a Hulth\'en one), 
which allow analytic calculations, providing a better 
insight on some of the features that the approach evidences.

\section{Expressions of form factors in ``point-form''}

\noindent
The expression of the pion form factor can be defined quite generally as: 
\begin{equation}
\sqrt{2E_f\;2E_i} \; \left<f|J^{\mu}|i\right> = F_1(q^2)\,(P^{\mu}_f+P^{\mu}_i). 
\label{2a}
\end{equation}
By identifying the matrix element of the current with its expression in terms 
of wave functions, an expression for the form factor $F_1(q^2)$ can be obtained. 
We first remind below results obtained for spin-0 constituents and then present 
expressions for the spin-1/2 case.

\subsection{Spin-0 constituent case}

\noindent
The matrix element relative to the ``point-form'' form factor 
for a system of spin-less particles may be written:
\begin{eqnarray}
\nonumber
\lefteqn{
\sqrt{2E_f\,2E_i} \; \left<f|J^{\mu}|i\right> = \sqrt{2M_f\,2M_i} \;  
\frac{1}{(2\pi)^3 } \int d^4p \, d^4p_f \,  d^4p_i \, d\eta_f \, d\eta_i  \,
\theta( \lambda_f \cdot p_f) \; \theta(\lambda_f \cdot p) \;
\theta(\lambda_i \cdot p)  \; \theta(\lambda_i \cdot p_i) }  \\
&&  \times  
\delta(p^2-m^2) \;  \delta(p^2_f-m^2) \; \delta(p^2_i-m^2) \;
\delta^4(p_f+p-\lambda_f \eta_f) \; \delta^4(p_i+p-\lambda_i \eta_i) 
\; \phi_f \left( (\frac{p_f-p}{2})^2 \right)  
\; \phi_i \left( (\frac{p_i-p}{2})^2 \right) 
\nonumber \\ 
&&   \hspace{5cm} \times
\left[\sqrt{(p_f+p)^2 \, (p_i+p)^2 } \;\;\;(p_f^{\mu}+ p_i^{\mu}) \right]. 
\label{2b}
\end{eqnarray}
Notations have been explained in Ref.~\cite{Desplanques:2001zw}. 
Let us mention here that the four-vectors $\lambda$ represent the velocities
of the system in the initial and final states, 
$\lambda^{\mu}_{i,f}=P^{\mu}_{i,f}/M$.  

\begin{figure}[htb]
\begin{center}
\includegraphics[width=20em]{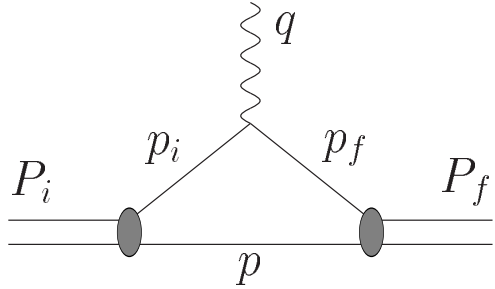}
\end{center}
\caption{Graphical representation of a virtual photon absorption on a pion  
together with kinematical definitions.}\label{fig:1}
\end{figure}

The kinematics is otherwise given in Fig.~\ref{fig:1}. The $\eta$ variables 
have been introduced to make the covariance explicit.
After performing some of the integrals, the matrix 
element of the current may be written:
\begin{eqnarray}
\sqrt{2E_f\,2E_i} \; \left<f|J^{\mu}|i\right> &=& \sqrt{2M_f\,2M_i} \;  
\frac{1}{(2\pi)^3 } \int \frac{d\vec{p}}{2\,e_p} \;
\frac{\phi_f \left((\frac{p_f-p}{2})^2\right)  \;  
\phi_i \left( (\frac{p_i-p}{2})^2 \right) }{(2\, \lambda_i \cdot p_i)\;
(2\, \lambda_f \cdot p_f)}
\nonumber \\ && \qquad \times
\left[ \sqrt{(p_f+p)^2 \, (p_i+p)^2 }\;\;\;(p_f^{\mu}+ p_i^{\mu})\right] . 
\label{2c}
\end{eqnarray}
In this expression, it is understood that the arguments $p_i^{\mu}$ 
and $p_f^{\mu}$ must be replaced by their value that stem from the 
$\delta(\cdots)$  functions appearing in Eq.~(\ref{2b}), which after 
integrating over the $\eta$ variables imply:
\begin{equation}
\vec{p}_i+\vec{p} = \frac{\vec{P}_i}{E_i}\;(e_i+e_p), \qquad\qquad
\vec{p}_f+\vec{p} = \frac{\vec{P}_f}{E_f}\;(e_f+e_p).
\label{2d}
\end{equation}
These relations characterize boost transformations in the ``point-form'' 
implementation proposed in Ref.~\cite{Klink:1998}. They differ from those used 
in instant- or front-form approaches.

In the Breit frame, the above expressions allow one to write 
the form factor, $F_1(q^2)$, as:
\begin{equation}
F_1(q^2)= \frac{1+v^2}{\sqrt{1-v^2}} \int \frac{d \vec{p}}{(2\pi)^3} 
\;\phi_f(\vec{p}_{tf}) \;\phi_i(\vec{p}_{ti}),
\label{2e}
\end{equation}
where $Q^2=-q^2$, $v^2=Q^2/(4\,M^2_{\pi}+Q^2)$ and
$\vec{p}_{ti,f}^{\,2}=p_x^2+p_y^2+(p_z\pm v\,e_p)^2/(1-v^2)$. 
It can be checked that 
this last expression of the form factor, or Eq.~(\ref{2c}) after an 
appropriate change of variable, leads to the relation:
\begin{equation}
F_1(0)=  \int \frac{d \vec{k}}{(2\pi)^3} 
\;\phi^2(\vec{k})=1,
\label{2f}
\end{equation}
in complete agreement with the charge of the system on the one hand, 
the ortho-normalization of the solutions of a mass operator on the 
other hand (no extra k-dependent factor is introduced in the integrand).  

The expression of Eq.~(\ref{2b}) may not be unique however. 
Due to the somewhat arbitrary 
appearance of the mass term, $M$, at the r.h.s., which should come out 
dynamically, an improved expression was proposed in Ref.~\cite{Amghar:2002jx}. 
Apart from the fact that it does not allow one to get analytic  
formulas, it does not add anything to the present discussion. 

\subsection{Spin-1/2 constituent case}

\noindent
In order to calculate the pion form factor, one has to take into account the
spinorial properties of the constituent quarks as well as the pseudo-scalar
nature of the pion. A minimal covariant factor accounting for these 
features involves the following trace of $\gamma$ matrices:
\begin{eqnarray}
I^{\mu}&\equiv&4 \;{\rm Tr}\left\{ \gamma_5\; \frac{(m+\gamma \cdot p_i)}{2}  
\;\gamma^{\mu}\, \frac{(m+\gamma \cdot p_f)}{2}\;\gamma_5\; 
\frac{(m-\gamma \cdot p)}{2}\right\}
 \nonumber \\
 &=& 2\; \left[p_i^{\mu}\;(p_f \cdot p)+ p_f^{\mu}\;(p_i \cdot p)- 
p^{\mu} \;(p_i \cdot p_f) + m^2\;(p_i^{\mu}+p_f^{\mu}+p^{\mu})\right].
\end{eqnarray}
At zero momentum transfer and when all particles are put on mass shell, 
this quantity becomes identical 
to $\sqrt{(p_f+p)^2 \, (p_i+p)^2 }\,(p_f^{\mu}+ p_i^{\mu})$, which is just the
factor that appears in the expression of the matrix element for scalar
constituents, see Eq.~(\ref{2b}). To get an expression for the 
pion form factor consistent with the normalization adopted for the solutions of 
the mass operator, it is therefore sufficient
to replace this factor for scalar particles by the above expression $I^{\mu}$. 
One thus gets the following matrix element for the $J^{\mu}$ current: 
\begin{eqnarray}
\nonumber
\lefteqn{
\sqrt{2E_f\,2E_i} \; \left<f|J^{\mu}|i\right> = \sqrt{2M_f\,2M_i} \;  
\frac{1}{(2\pi)^3 } \int d^4p \, d^4p_f \,  d^4p_i \, d\eta_f \, d\eta_i  
\; \theta( \lambda_f \cdot p_f) \; \theta(\lambda_f \cdot p) 
\; \theta( \lambda_i \cdot p)   \; \theta(\lambda_i \cdot p_i) }  \\
&&   \times \; 
\delta(p^2-m^2) \;  \delta(p^2_f-m^2) \; \delta(p^2_i-m^2) 
 \delta^4(p_f+p-\lambda_f \eta_f) \;\delta^4(p_i+p-\lambda_i \eta_i) \;
\; \phi_f \left( (\frac{p_f-p}{2})^2 \right)    
\; \phi_i \left( (\frac{p_i-p}{2})^2 \right) 
\nonumber \\ 
&&  \qquad\qquad  \times \,
2\,\left[p_i^{\mu}\;(p_f \cdot p)+ p_f^{\mu}\;(p_i \cdot p)- 
p^{\mu} \;(p_i \cdot p_f) + m^2\;(p_i^{\mu}+p_f^{\mu}+p^{\mu})\right].
\label{2i}
\end{eqnarray}
In the Breit frame, the matrix element of Eq.~(\ref{2i}) 
leads to the expression for the form factor:
\begin{equation}
F_1(Q^2)= \sqrt{1-v^2}
\int \frac{d \vec{p}}{(2\pi)^3} 
\;\phi_f(\vec{p}_{tf}) \;\phi_i(\vec{p}_{ti}).
\label{2j}
\end{equation}
It can be checked again that this last expression of the form factor is
in complete agreement with the charge of the system and
the ortho-normalization of the solutions of a mass operator.

It is noticed that the form factor of Eq.~(\ref{2j}) differs from the one 
obtained with scalar constituent particles by 
a factor $(1-v^2)/(1+v^2)$, see Eq.~(\ref{2e}). 
This last factor is close to $1-Q^2/ (8\,m^2)$ for small momentum transfers
and small binding energies ($M\simeq 2\,m$). It may be identified with the 
Darwin-Foldy term sometimes introduced in the calculation of form factors 
involving spin-1/2 constituents. A similar factor can be found 
from Ref.~\cite{Chung:1988mu} as a result of Melosh rotations where, however, 
a quantity $2m$ appears in place of the pion mass $M_{\pi}$ 
in the definition of $v$.

\section{Analytic results for form factors}

\noindent
In a few cases, analytic expressions of form factors given by Eq.~(\ref{2j}) 
can be obtained. This is the case for a Gaussian 
wave function, that has been used widely in the field 
(Refs.~\cite{Isgur:1984jm,Chung:1988mu,Krutov:2002nu,Allen:1998hb}),
and a Hulth\'en one\footnote{The relevance of power-law wave functions with
respect to hadron form factors was especially noted in 
Ref.~\protect\cite{Cardarelli:1995dc}.}. The latter one is of interest because 
it leads to the correct asymptotic power law 
for the form factor in the case of scalar constituents.
These wave functions with the appropriate normalizations 
are given by:
\begin{equation}
\phi^G(\vec{k}) = \frac{(4\pi)^{3/4}}{b^{3/2}}\;
\exp \left[-k^2/(2\,b^2)\right],\qquad\qquad
\phi^H(\vec{k}) = \sqrt{4\,\pi} \; \frac{\sqrt{2\, \alpha\; \beta\; 
(\alpha+ \beta)^3}}{(\alpha^2+k^2)\; (\beta^2+k^2)}.
\label{3a}
\end{equation}
The parameters may be chosen as free parameters 
or fitted to the matter radius. For the Hulth\'en wave function, 
$\alpha$ is in principle related to the binding 
energy by $\alpha^2=m^2-M^2/4$, only $\beta$ is really a free parameter.  
With these wave functions the integral in Eq.~(\ref{2j}) may be calculated
analytically. One gets in the Gaussian case:
\begin{equation}
F_1^G(Q^2)=
\sqrt{1-v^2} \;\int d \vec{p} \;
\left(\frac{1}{b^{3/2}\; \pi^{3/4}}\right)^2 \;  
\exp\left[ -\frac{1}{b^2} 
 \left( p^2_{\perp}+ \frac{ p^2_{\parallel}+v^2\;e^2_p}{1-v^2} \right) \right] =
\frac{(1-v^2)^2}{ \sqrt{1+v^2} }\;
\; \exp \left[- \frac{v^2}{1-v^2} \;\frac{m^2}{b^2}\right],
\label{3b}
\end{equation}
and the associated charge radius and asymptotic behavior are given by:
\begin{equation}
\left<r^2\right>^G\equiv -6 \;\frac{dF_1^G(Q^2)}{dQ^2} =\frac{6}{4\,M_{\pi}^2}\;
\left( \frac{5}{2} + \frac{m^2}{b^2} \right),\qquad
\left.F_1^G(Q^2)\right|_{Q^2\rightarrow \infty} 
=\frac{1}{\sqrt{2}}\;\frac{16\,M_{\pi}^4}{Q^4} \;
\;\exp \left[- \frac{Q^2}{16\,b^2} \;\frac{4\,m^2}{M_{\pi}^2}\right].
\label{3c}
\end{equation}
In the case of the Hulth\'en wave function, the form factor reads:
\begin{eqnarray}
\nonumber 
F_1^H(Q^2)&=&\sqrt{1-v^2} \,\int \frac{d \vec{p}}{(2\pi)^3} \;
8\pi \alpha \beta\; (\alpha+ \beta)^3
\nonumber \\ &&  \times 
\frac{1}{ \left( \alpha^2 + p^2_{\perp} +                        
(\frac{p_{\parallel}-v\,e_p}{\sqrt{1-v^2}})^2 \right)
\;     \left( \beta^2 + p^2_{\perp} +                        
(\frac{p_{\parallel}-v\,e_p}{\sqrt{1-v^2}})^2 \right)  
\left( \alpha^2 + p^2_{\perp} + 
(\frac{p_{\parallel}+v\,e_p}{\sqrt{1-v^2}})^2 \right)
\;        \left( \beta^2 + p^2_{\perp} + 
(\frac{p_{\parallel}+v\,e_p}{\sqrt{1-v^2}})^2 \right) }
\nonumber \\ &=&
\frac{(1-v^2)^2}{v}\; \; \frac{\alpha\; \beta\; 
(\alpha+ \beta)}{(\alpha- \beta)^2} 
\left( \frac{\arctan[v\sqrt{m^2-\alpha^2}\,/\alpha]}{\sqrt{m^2-\alpha^2}}
-\frac{{\rm 
arctan}[2\,v\sqrt{m^2-\alpha^2}\,/(\alpha+\beta-v^2(\beta-\alpha))]
}{\sqrt{m^2-\alpha^2}}
\right.  \nonumber \\ &&  \qquad \qquad \left. + \,
 \frac{\arctan[v\,\sqrt{m^2-\beta^2}\,/\beta]}{\sqrt{m^2-\beta^2}}
- \frac{{\rm 
arctan}[2\,v\sqrt{m^2-\beta^2}\,/(\alpha+\beta+v^2(\beta-\alpha))]
}{\sqrt{m^2-\beta^2}}  \right).
\label{3d}
\end{eqnarray}
The charge radius and high $Q^2$  properties associated to this
form factor are given by:
\begin{eqnarray}
\left<r^2\right>^H&=&\frac{6}{4\,M_{\pi}^2}\;
\left( 2
+ \frac{\alpha\;(7\,\beta^2+4\,\beta\,\alpha+\alpha^2)}{
3\,\beta^2\;(\beta+\alpha)} +
\frac{m^2-\alpha^2}{3}\;\frac{ \beta^4 +  5\,\beta^3\,\alpha+
12\,\beta^2\,\alpha^2 + 5\,\beta\,\alpha^3 + \alpha^4}{
\beta^2\,\alpha^2\;(\beta+\alpha)^2}\right) \nonumber \\
&& \qquad
\stackrel{\beta\rightarrow\infty}{\longrightarrow}
\frac{3}{M_{\pi}^2}+\frac{1}{8\, \alpha^2},
\nonumber \\
\left.F_1^H(Q^2)\right|_{Q^2\rightarrow \infty} 
&=&\frac{64\,M_{\pi}^8\; \alpha\,\beta\;(\beta+\alpha)^2}{Q^8\;(m^4+\ldots)}
\quad\stackrel{\beta\rightarrow\infty}{\longrightarrow}\quad
\frac{16\,M_{\pi}^4}{Q^4} \; \alpha
\; \frac{\arctan[\sqrt{m^2-\alpha^2}\;/\alpha]}{\sqrt{m^2-\alpha^2}},
\label{3e}
\end{eqnarray}
where the dots in the last expression represent contributions that can be 
neglected for finite values of $\beta$. 
An interesting limit is the Coulombian case $(\beta=\alpha)$, considered 
in Ref.~\cite{Desplanques:2001zw}  for scalar constituents. 
The expressions in Eq.~(\ref{3e}) then simplify to read:
\begin{equation}
\left<r^2\right>^C=\frac{6}{M_{\pi}^2}+\frac{3}{4\, \alpha^2}, \qquad\qquad
\left.F_1^C(Q^2)\right|_{Q^2\rightarrow \infty}
=\frac{256\,M_{\pi}^8\; \alpha^4}{Q^8\; m^4} .
\label{3g}
\end{equation}
%

\section{Discussion}

\subsection{Charge radius}

\noindent
As it can be observed from Eqs.~(\ref{3c}) and~(\ref{3e}), the squared 
charge radius scales like the inverse squared pion mass, confirming 
the hand-waving argument given in the introduction and, at the same time, 
the consequence that the charge radius increases with increasing
binding energy!

The second observation is that the squared charge radius 
has a minimum value, as a function of the open parameters, 
that depends on the pion mass. It slightly 
differs with the model ($15/(4\,M_{\pi}^2)$ for the Gaussian wave function 
and $3/M_{\pi}^2$ for the Hulth\'en one). 
However, even this minimal values, ($7.5\,{\rm fm}^2$ 
and $6\,{\rm fm}^2$) exceed by far the experimental one, 
$0.43\,{\rm fm}^2$~\cite{Amendolia:1986wj}. 
Independently, the contribution of the bare matter radius 
of the pion to these values is not small. This part, $3/(8b^2)$, 
is enhanced by 
$(2\,m/M_{\pi})^2$, which amounts to a factor $8$ - $18$ for quark masses in the
range of $0.2$ - $0.3$ GeV. For a squared matter radius of  
$0.05\,{\rm fm}^2$, of the order of what is generally predicted by 
quark models, the corresponding squared charge radius is already 
of the order of the experimental one. The existence of a minimum 
value largely exceeding the experimental one has the striking
consequence that there is no way to fit the parameters of the model 
to reproduce the experimental pion form factor.

\subsection{Asymptotic behavior}

\noindent
The asymptotic form factor for a system made of 
spin-1/2 constituents is expected to scale like
$Q^{-2}$~\cite{Brodsky:1975vy,Lepage:1979zb}.
In addition, the coefficient of the $Q^{-2}$ factor should be
proportional to the strength of the short range quark-quark interaction 
(up to log terms) and to the square of the wave function at the origin. 
These expectations are independent of any assumption on the mass of the 
constituents or on the size of the system. 

Evidently, the Gaussian model cannot reproduce 
the correct power law behavior of form factors in general. Only in the zero-size 
limit, when $b^2\rightarrow\infty$, one obtains 
$F_1 \propto (M_{\pi}^2/Q^2)^2$. Apart from 
the fact that the power of $Q$ is not the right one, the overall coefficient 
does not scale like the strength of the quark-quark interaction. 
This illustrates the statement that reproducing the QCD power-law behavior 
with Gaussian wave functions and standard currents is most surely 
the consequence of a mistake in the theoretical approach employed 
to get the result. 

Not surprisingly, form factors calculated with Hulth\'en wave 
functions provide a power law behavior ($Q^{-8}$), which, however, differs
from what is expected. For a large part, this is in relation 
with the observation made in Ref.~\cite{Allen:2000ge} that the transferred 
momentum to the struck particle in the ``point-form'' approach is  
$Q^2\;[1+Q^2/(4\,M^2)]$ rather than $Q^2$. 
A better result is obtained in the limit 
$\beta \rightarrow \infty$, in which case the wave function, 
Eq.~(\ref{3a}), scales like $k^{-2}$ instead of  $k^{-4}$. 
This limit however corresponds to a zero-range force, which 
is difficult to imagine for an effect dominated by one-gluon 
exchange. 

Apart from that, one can check that the expression 
of the form factor, Eq.~(\ref{3d}), has at least some of the 
appropriate properties which can be easily checked in the 
Coulombian case, Eq.~(\ref{3g}). The coefficient in the expression for the
asymptotic form factor, $\alpha^4$, 
splits into a factor $(\alpha^{3/2})^2$, representing 
the square of the wave function at the origin, and a factor $\alpha$ 
which involves the strength of the short-range force 
($\alpha=m\,\alpha_s/2$). The same observation applies
to the Hulth\'en wave function.

Concerning the asymptotic behavior itself, one generally refers to the $Q^{-2}$
power law from QCD but it seems that this is not what one should expect from a
relativistic quantum mechanics calculation based on a single-particle current.
Calculations in the light front approach~\cite{Carbonell:1998rj}, in the instant
form and in the present work (after disentangling the effect of the modified
momentum transfer discussed above), rather indicate a $Q^{-4}$ asymptotic 
behavior, like for scalar constituents. One has to rely then on two-body 
currents to get the correct $Q^{-2}$ behavior. It can be checked that this 
behavior is obtained without any problem from a single particle current in the 
case of a scalar probe. This indicates that the different power law obtained for
spin-1/2 constituents has its origin in the specific form of the coupling 
between quarks and photons (see below). 

\subsection{Discrepancy with an earlier calculation}

\noindent
Results presented above significantly differ from those obtained in 
Ref.~\cite{Allen:1998hb}, where the charge radius was correctly reproduced. 
For an unknown reason\footnote{This factor was claimed to be
necessary to get the non-relativistic limit right but the sense of this
statement is not transparent and the recipe was not used anymore in
later works.},
a factor $M_{\pi}/(2\,m)$ was introduced in the relation between
the momentum of the pion in the Breit frame and the momentum transfer, 
$\vec{P}_{i,f}=\pm (\vec{Q}/2) \times M_{\pi}/(2\,m)$ 
(see Eq.~(14) of Ref.~\cite{Allen:1998hb}). The velocity $v^2$, 
which is the relevant parameter in the present ``point-form'' approach,  
thus becomes:
\begin{equation}
v^2=\frac{Q^2}{4\,M_{\pi}^2+Q^2} 
 \qquad \longrightarrow  \qquad
v^2=\frac{Q^2}{16\,m^2+Q^2}.
\label{4b} 
\end{equation}
With the last expression, all problems related to the 
limit $M_{\pi} \rightarrow 0$ vanish. There is no more lower bound
for the squared charge radius that prevents one from making a fit 
of the form factor to experimental values. While the above modification 
brings the pion form factor closer to those obtained in other forms, 
there is no justification for this change within the ``point-form'' 
approach. Evidently, the more reasonable character of the results obtained 
with this correction suggests that there may be some truth in it 
\cite{Desplanques:2001ze} but no tentative explanation was 
presented by the authors. 

The other difference concerns the asymptotic form factor obtained 
with a Gaussian wave function. We gave arguments to discard 
the use of such wave functions for looking at the asymptotic 
behavior of form factors. It however remains to explain 
why, in the zero-size limit, a $Q^{-2}$ behavior was obtained in 
Ref.~\cite{Allen:1998hb} while we get $Q^{-4}$. 
This is most likely due to the matrix element of the current,
which in the Breit frame reads:
\begin{equation}
\bar{u} \; \gamma^0 \; u \simeq  
1- \frac{(\vec{\sigma} \cdot \frac{\vec{Q}}{2})^2 }{(m+e_{Q/2})^2} 
\simeq \frac{2\,m }{m+e_{Q/2}}.
\label{4a}
\end{equation}
Our expression for the form factor implicitly accounts for the 
full structure of the spinors, thus evidencing some dependence 
on the momentum transfer, as can be seen 
from Eq.~(\ref{4a}). This tends to reduce the value 
of the matrix element of the charge density. 
In  Ref.~\cite{Allen:1998hb}, this matrix element was assumed to be a constant:
$\left<p_1' \;  \sigma_1'\right|J_1^0(0)\left|p_1 \; \sigma_1\right> = 
e_1\; \delta_{\sigma_1' \; \sigma_1}$,
which differs from Eq.~(\ref{4a}) by a factor $(m+e_{Q/2})/(2\,m)$. 
It can be checked that a full account of Eq.~(\ref{4a}), together with the 
redefinition of $v^2$ as given in Eq.~(\ref{4b}) and the modified relation 
between the momentum transfer and the
momentum of the struck particle in the ``point-form'' approach, provides 
an extra factor $(1-v^2)$. Dividing Eq.~(\ref{3b}) by this quantity, the form 
factor would read:  
\begin{equation}
F_1(Q^2)= \frac{1-v^2}{\sqrt{1+v^2}},
\label{4d} 
\end{equation}
which is close to the results of Ref.~\cite{Allen:1998hb} with
a quark mass fitted to the experimental charge radius ($m=0.22\;$GeV). 

\subsection{Numerical estimates of form factors}

\begin{figure}[tb]
\begin{center}
\mbox{\includegraphics[width=0.48\textwidth]{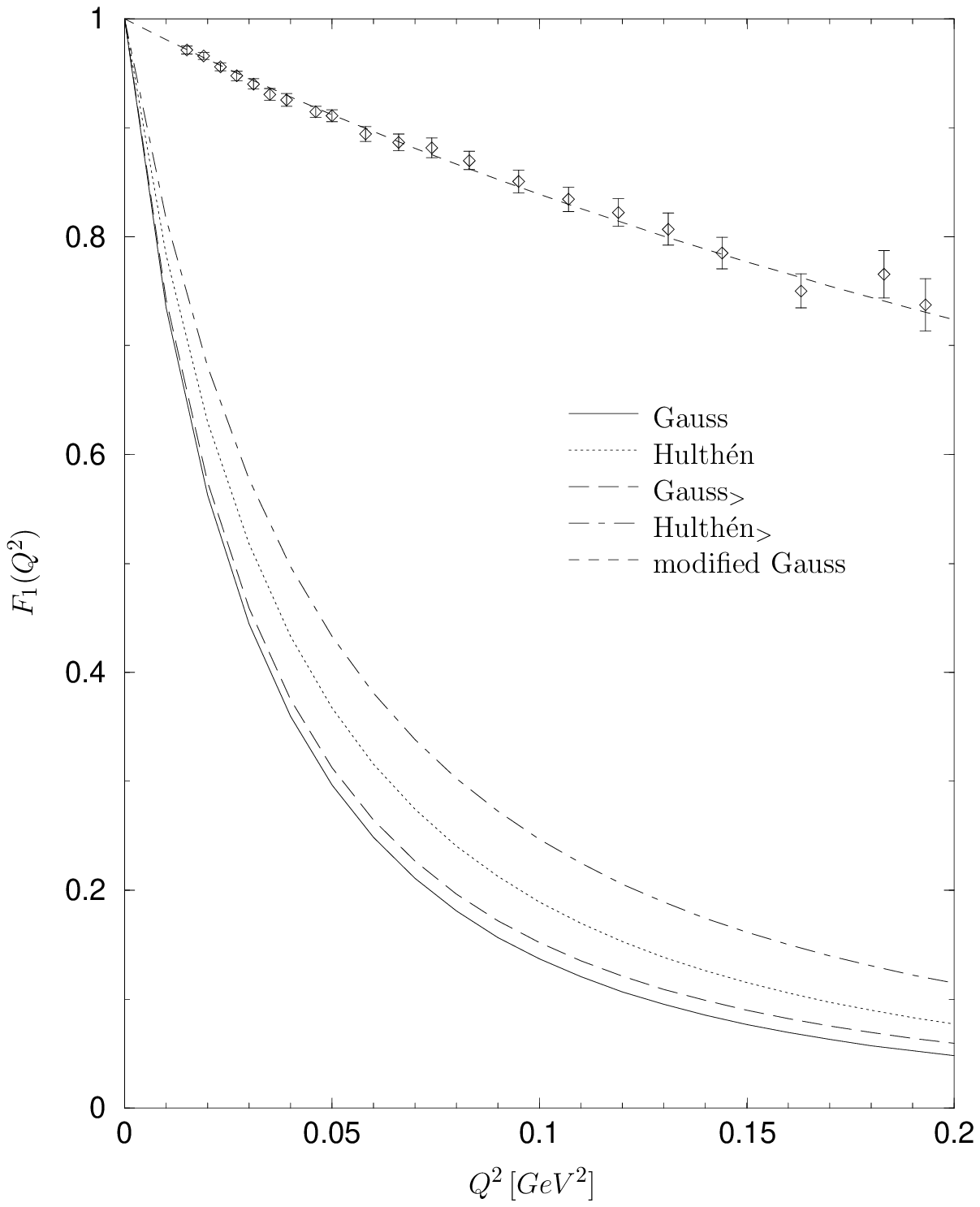}
\includegraphics[width=0.48\textwidth]{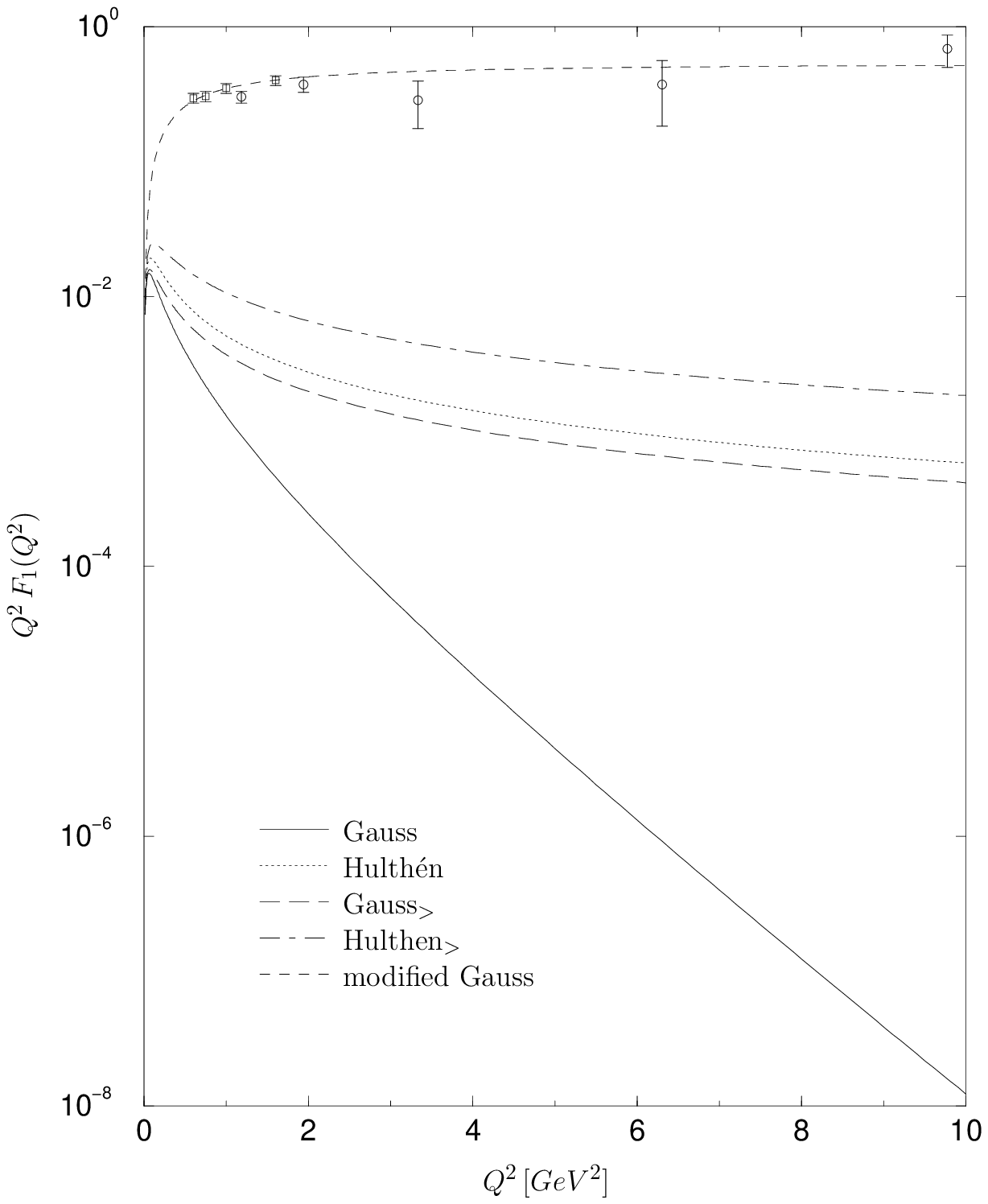}}
\end{center}
\caption{Pion form factor calculated with different wave functions. The left
hand graph shows the low-$Q^2$ behavior in linear scale, the graph on the 
right hand side shows the form factors, multiplied by $Q^2$, in logarithmic
scale up to $Q^2=10$ GeV$^2$. The different curves correspond to 
Eq.~(\protect\ref{3b}) (solid), Eq.~(\protect\ref{3d}) with
$\beta\rightarrow\infty$ (dotted),
Eq.~(\protect\ref{4f}) (long- and dot-dashed) and Eq.~(\protect\ref{4d}) 
(dashed), respectively. Experimental data are from 
Refs.~\protect\cite{Bebek:1978pe,Amendolia:1986wj,Volmer:2000ek}.
}\label{fig:2}
\end{figure}

\noindent
In Fig.~\ref{fig:2} we provide some numerical estimates for form factors 
calculated 
with Gaussian and Hulth\'en wave functions. This is done for what 
we consider the most preferable case with respect to a reasonable 
choice of parameters and a favorable
comparison to experiment. For the Gaussian wave function, the 
parameter $b$ is chosen such as to get a matter radius of $0.025\,{\rm fm}^2$, 
which is at the lower bound of predictions~\cite{Carlson:1983rw}, giving 
$b^2=0.6\;{\rm GeV}^2$. The quark mass is taken as $m=0.22\;$GeV. For the 
Hulth\'en wave function, we take the limit $\beta \rightarrow \infty$ and the 
parameter $\alpha$ is fixed to give the same matter radius as in the Gaussian 
case, giving $\alpha^2=0.2\;{\rm GeV}^2$. 
The quark mass here is chosen such that $m^2-\alpha^2=M_{\pi}^2/4$. 

In addition, Fig.~\ref{fig:2} contains the form factor given by Eq.~(\ref{4d}), 
with the redefinition of $v^2$ as discussed before, see Eq.~(\ref{4b}). With a
value of $m=0.22\;$GeV, one obtains a good representation of both 
experimental data and  results obtained in  Ref.~\cite{Allen:1998hb} 
for similar conditions. Evidently, the Gaussian and Hulth\'en form factors 
miss the experimental values by a large factor. 
In the very extreme but unphysical limit of a point-like pion, 
the following upper limits are obtained:
\begin{equation}
F_1^G(Q^2)_{>}= \frac{(1-v^2)^2}{\sqrt{1+v^2}}, \qquad \qquad
F_1^H(Q^2)_{>}=\frac{(1-v^2)^2}{2v} \; \log \frac{1+v}{1-v}.
\label{4f} 
\end{equation}
These ones, also reported in Fig.~\ref{fig:2}, are still far 
below the experimental values.

\section{Conclusion}

\noindent
In a previous work~\cite{Desplanques:2001zw,Amghar:2002jx}, 
``point-form'' form factors were calculated and found 
to evidence a large discrepancy, both at low and high $Q^2$, with what could 
be considered an ``exact'' calculation. The system 
under consideration was academic however. In the present work, 
we applied the same approach  to the pion form factor whose low $Q^2$ 
behavior (related to the charge radius) and high $Q^2$ behavior 
are determined sufficiently well from experiment to make reliable  
statements. Results we obtained, based on  a single-particle current, 
evidence quite similar features: too fast fall off at both low 
and high $Q^2$. In the first case, they point to a squared charge radius  
more than one order of magnitude larger than the measured one. 
In the  second case, the asymptotic behavior misses by many powers 
of $Q$ the $1/Q^2$ power law expected from QCD.

For our purpose, we used simple wave functions which offer the 
advantage of providing analytic results. Better wave functions 
could be used. However, the comparison with results from
Refs.~\cite{Desplanques:2001zw,Amghar:2002jx} 
does not show qualitative differences as far as 
the most striking features are concerned. At low $Q^2$,
results obtained with Gaussian and Hulth\'en wave functions are very similar.
Both point to an uncompressible value of the squared 
charge radius of the order of $6\,{\rm fm}^2$, 
an order of magnitude larger than the experimental value. 
This prevents one from making any 
fit to the measured form factor. At high $Q^2$, the model 
dependence may be more important but, apart from the fact that 
the correct power law behavior is missed, the most favorable cases,
such as the zero-size limit, are physically irrelevant. 
In all  cases we considered, the ``point-form'' form factors are 
far below experimental data. 

Present  results differ strikingly from those obtained in 
Ref.~\cite{Allen:1998hb}. 
For the largest part, this is due to a different
relation between the pion Breit-frame momentum and the 
momentum transfer which was used by these authors, allowing 
them to escape the constraint of a minimum value 
of the squared charge radius determined by the pion mass. However,
this relation has no theoretical support within the 
``point-form'' approach. 

On the basis of a theoretical model, it was concluded  
in a previous work that the present implementation 
of the ``point-form'' approach together with a single-particle 
current requires major improvements~\cite{Desplanques:2001zw}. 
The present results for a physical quantity, the pion form factor, 
lead to the same conclusion. Large contributions of two-body 
currents are needed~\cite{Desplanques:2003}. 
An alternative would be to improve the present implementation of the 
``point-form'' approach~\cite{Theussl:2003fs}. 

\begin{acknowledgments} 
We are grateful to W. Klink for confirming some points of his work 
that we reconsidered here.
This work has been supported by the EC-IHP Network ESOP under contract 
HPRN-CT-2000-00130 and  MCYT (Spain) under contract BFM2001-3563-C02-01.
\end{acknowledgments}



\end{document}